\documentstyle[epsfig]{aipproc}

\begin{document}
\title{ A Simple BATSE Measure \\
of GRB Duty Cycle }

\author{Jon Hakkila$^*$, Robert D. Preece$^{\dagger}$, \& Geoffrey N. Pendleton$^{\dagger}$}
\address{$^*$Minnesota State University, Mankato, Minnesota 56001\\
$^{\dagger}$University of Alabama, Huntsville, Alabama 35812}

\maketitle

\begin{abstract}
We introduce a definition of gamma-ray burst (GRB) duty cycle 
that describes the GRB's efficiency as an emitter; it is the GRB's 
average flux relative to the peak flux. This GRB duty cycle is 
easily described in terms of measured BATSE parameters; it is 
essentially fluence divided by the quantity peak flux times duration.

Since fluence and duration are two of the three defining 
characteristics of the GRB classes identified by statistical 
clustering techniques (the other is spectral hardness), duty cycle 
is a potentially valuable probe for studying properties of these classes.
\end{abstract}

\section*{Introduction}

The term {\it duty cycle} in astrophysics is defined as ``the 
fraction of time a pulsed beam is on'' \cite{hopkins80}. This term is 
more appropriate in describing periodic emitters such as pulsars than 
it is for non-periodic, one-time emitters such as gamma-ray bursts (GRBs). 
GRB emission consists of pulses varying in intensity over a burst's 
duration, and is thus more conducive to a definition recognizing the 
continuous nature of GRB emission than to one limited as either ``on''
or ``off.'' A more appropriate definition of GRB duty cycle should 
describe the effectiveness of a GRB as an emitter during the time 
that it emits. We therefore define the {\it GRB duty cycle} as the 
average flux relative to its peak flux. This duty cycle definition 
can be described in terms of measured BATSE parameters; it is 
essentially fluence divided by the quantity peak flux times duration.

Fluence and duration are two of the three defining characteristics 
of the three GRB classes identified by statistical clustering 
techniques \cite{mukherjee98}. Spectral hardness is the third.
Fluence (time-integrated flux) incorporates information about duration 
in its definition. Overlapping information appears 
to be contained in fluence and duration \cite{bagoly98}. For
this reason, duty cycle (as defined here) is a potentially valuable 
probe for studying properties of the three GRB classes.

Properties of the three GRB classes as determined from statistical 
clustering techniques \cite{mukherjee98} are demonstrated in Table 
\ref{table1}.

\begin{table}[ht!]
\caption{Statistical clustering classes, from 3B GRBs.}

\label{table1}
\begin{tabular}{lccc}
 Attributes& Class 1 (Long)& Class 2 (Short)& Class 3 (Intermediate) \\
\tableline
T90: & long & short & intermediate \\
Fluence: & large & small & intermediate \\
Hardness: & intermediate & hard & soft \\
\end{tabular}
\end{table}

\section*{Duty Cycle Definition}

We define the duty cycle (DC) in terms of BATSE parameters as:

\begin{equation}
DC= \frac{S_{23}}{A \cdot P_{64} \cdot T_{90}}
\end{equation}

Here, S$_{23}$ is the channel 2+3 fluence (time-integrated flux between 
50 and 300 keV), T$_{90}$ is the duration spanning 90\% of the GRB 
emission, and P$_{64}$ is the 64 ms peak flux. $A$ is a constant for 
converting photon counts to energy (assuming a diagonal detector 
response matrix as a first-order approximation).

The peak flux used in this calculation must be measured on the shortest 
available timescale (64 ms) in order to avoid arbitrarily smoothing 
out the maximum value of the peak flux. A peak flux underestimate 
produces a corresponding duty cycle overestimate.

GRBs with T$_{90}$ durations less than 64 ms have had their T$_{90}$ 
values set to 64 ms, so that their durations are not given a different 
temporal resolution than that of their peak flux measure.

\section*{Analysis of GRB Classes}

Our database consists of non-overwriting GRBs in the BATSE 4Br Catalog \cite{paciesas99} triggering on 1024 ms peak flux in the 50 to 300 keV 
range with the trigger threshold set $5.5 \sigma$ above background. 
These criteria prevent trigger biases \cite{meegan99} from influencing 
our conclusions. We have also removed GRBs with large relative 
measurement errors in each of the four-channel fluences, T$_{90}$, and 
P$_{64}$ so that measurement error does not bias our conclusions.

The GRBs have been assigned to a class using the supervised decision 
tree classifier C4.5 \cite{quinlan86}. The technique is described in 
more detail elsewhere \cite{hakkila99}. 

We obtain an average value of $A \approx 2.24 \times 10^{-7}$ ergs 
photon$^{-1}$ by integrating typical bright GRB spectrum 
(soft power law index $\alpha \approx -1$) over the 50 to 300 keV 
trigger energy range.

Many Class 2 (Short) GRBs have duty cycles $DC > 1.0$, as their harder 
spectra lead to an underestimate of A and to a corresponding 
overestimate of DC. We obtain a separate value of $A \approx 2.80 \times 
10^{-7}$ ergs photon$^{-1}$ for these GRBs (using $\alpha 
\approx 0$), and recalculate their duty cycles. No attempt is made to 
account for Class 3 (Intermediate) spectra, which have similar spectral components to those of faint Class 1 (Long) bursts \cite{hakkila99}.

\begin{figure}[ht!] 
\centerline{\epsfig{file=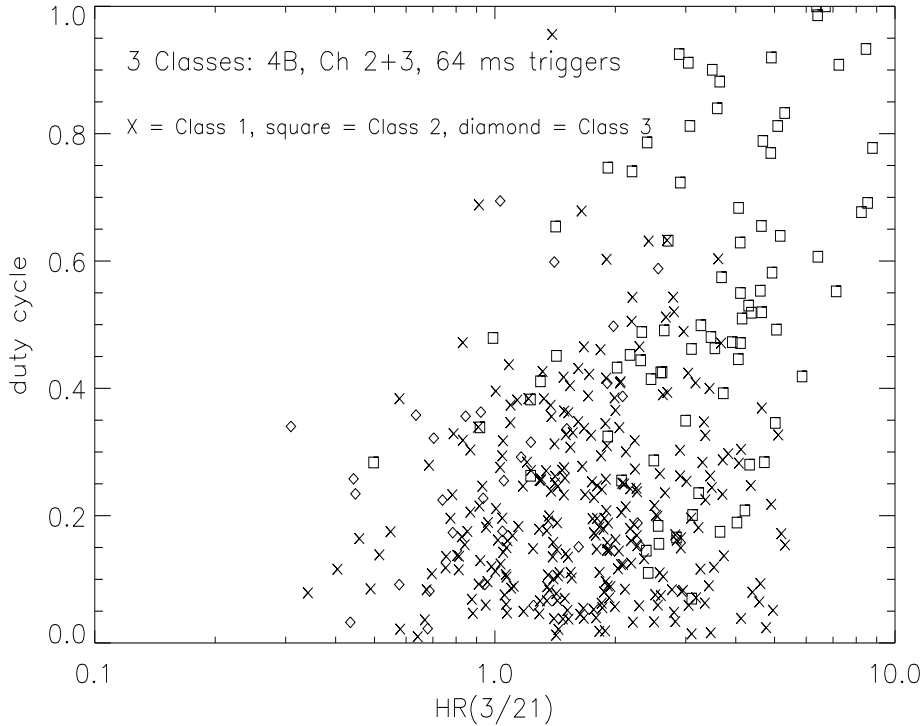,height=4.0in,width=5.0in}}
\vspace{10pt}
\caption{GRB Duty Cycle vs. hardness ratio HR321.}
\label{fig1}
\end{figure}

Figure \ref{fig1} is a plot of $DC$ vs. hardness ratio 
$HR321$ (100 to 300 keV fluence divided by 25 to 100 keV fluence). 
This hardness ratio represents the third delineating attribute of 
the three GRB classes.

We note the following characteristics of GRBs, and of the three GRB 
classes (based upon Figure \ref{fig1}), as they pertain to the duty cycle:
\begin{itemize}
\item There are no efficient, soft GRBs.
\item There are no inefficient, hard GRBs.
\item Class 2 (Short) GRBs are generally efficient, 
with duty cycles of $DC \geq 0.1$.
\item Class 1 (Long) GRBs are rarely efficient, 
with duty cycles of $DC \leq 0.7$. 
\item Class 3 (Intermediate) GRBs blend into the Class 1 (Long) 
GRBs in this plot.
\end{itemize}

\section*{Conclusions} 

The duty cycle measure as defined here is fairly effective and easy 
to calculate. The simplifying assumption of a diagonalized detector 
response matrix is not completely valid. A small correction factor is 
needed to account for excess high-energy photons from hard GRBs 
(primarily those belonging to Class 2). Nonetheless, this approach
allows preprocessed BATSE attributes to be incorporated directly into
the duty cycle calculation, without requiring the use of data in a
less processed form.

Despite the aforementioned problem, Class 2 (Short) is well delineated 
from Class 1 (Long) in a plot of duty cycle vs. HR321. Class 2 GRBs 
are harder, more efficient emitters than Class 1 GRBs.

Class 3 (Intermediate) does not appear to be distinct from Class 1 
(Long) on the basis of the duty cycle attribute. This result is in 
agreement with the findings of our artificial intelligence study 
\cite{hakkila99}.

\end{document}